\documentclass[apj]{emulateapj}
\usepackage{graphicx}

\shorttitle{Temporal variability of active region outflows}
\shortauthors{Ugarte-Urra \& Warren}

\begin{document}


\title{Temporal variability of active region outflows}

\author{Ignacio Ugarte-Urra}
\affil{College of Science, George Mason University. 4400 University Drive, Fairfax, VA 22030, USA.}
\author{Harry P. Warren}
\affil{Space Science Division, Code 7670, Naval
       Research Laboratory, Washington, DC 20375, USA.}


\begin{abstract}
Recent observations from the Extreme-ultraviolet Imaging Spectrometer (EIS) on
board {\it Hinode} have shown that low density areas on the periphery of
active regions are characterized by strong blue-shifts at 1 MK. These Doppler
shifts have been associated with outward propagating disturbances observed
with Extreme-ultraviolet and soft X-ray imagers. Since these instruments
can have broad temperature responses we investigate these intensity fluctuations 
using the monochromatic imaging capabilities of EIS and confirm their 1 MK nature.
We also find that the \ion{Fe}{12} 195.119 \AA\ blue shifted spectral profiles
at their footpoints exhibit transient blue wing enhancements on timescales
as short as the 5 minute cadence. We have also looked at the fan peripheral loops 
observed at 0.6 MK in \ion{Si}{7} 275.368 \AA\ in those regions and find
no sign of the recurrent outward propagating disturbances with velocities
of 40 -- 130 $\rm km\,s^{-1}$ seen in \ion{Fe}{12}. We do observe downward
trends (15 -- 20 $\rm km\,s^{-1}$) consistent with the characteristic red-shifts
measured at their footpoints. We, therefore, find no evidence that the 
structures at these two temperatures and the intensity fluctuations they exhibit
are related to one another.

\end{abstract}



\section{Introduction}
At the periphery of solar active regions, rooted in strong flux concentrations,
there are cool loops with temperatures of 1\,MK  \citep{schrijver1999} and 
under, with footpoints that often adopt fan-like geometries; the so-called 
``fan loops''. High cadence EUV movies of these loops 
reveal upward propagating motions with projected velocities of 50 -- 150 $\rm km\,s^{-1}$
\citep{berghmans1999,schrijver1999,sakao2007, mcintosh2009} that have been
interpreted as flows, but also waves \citep{demoortel2002b,demoortel2002,wang2009}. 
Spectroscopic analyses show that these loops have temperatures of 0.6 -- 1 MK 
and densities comparable to those of the core loops, and emit particularly 
strongly in \ion{Mg}{6}, \ion{Mg}{7}, \ion{Si}{7} and \ion{Fe}{8} lines
\citep{delzanna2003,delzanna2003b, young2007}, which have
an ionization equilibrium temperature $\le10^{5.8}$ K. Perhaps in 
contradiction to the upward propagating disturbances, the emission of these
lines is consistently red-shifted
\citep{winebarger2002,marsch2004,delzanna2008, warren2010b}. This is
also consistent with the downflowing plasma that has 
been observed off-limb in high cadence monochromatic imaging 
\citep[e.g.][]{ugarte-urra2009}. The fans can live for hours and
days \citep{schrijver1999}, but the individual structures evolve on
timescales of the order of tens of minutes.

Also at the periphery, there are low density areas at 1 -- 2 MK, that are 
hardly noticeable in emission when compared to the core loops. 
In fact, only recent spectral measurements from the Extreme-ultraviolet 
Imaging Spectrometer (EIS) on board {\it Hinode} have revealed that these areas are 
very distinct in Doppler maps for their characteristic strong blue-shifts and 
enhanced broadening of \ion{Fe}{12} -- \ion{Fe}{15} lines 
\citep{doschek2008,harra2008,delzanna2008}. The shifts in the spectral line 
profiles are up to 50 $\rm km\,s^{-1}$, but asymmetries in the blue wings
expose contributions from plasma at 100 -- 200 $\rm km\,s^{-1}$ 
\citep{mcintosh2009,bryans2010}, indicative of outflows that often persist for 
many days. These regions lie over or near magnetic flux concentrations of a
single polarity \citep{doschek2008}.

\citet{mcintosh2009} have argued that these persistent blue-shifts in the
outskirts of active regions are the spectral signature of the disturbances
observed by EUV imagers in the 1 MK loops and suggest that the upflows are 
related to spicule activity.

In this paper, we find short timescale variability in the blue wing
enhancements in the  \ion{Fe}{12} 195.119 \AA\ line, supportive of the 
transient nature of the disturbances precursors. We do not, however, find
any obvious relationship between the temporal response of the low 
density plasma at 1 MK and the evolution of the high density 0.6 MK loops.
We do observe disturbances propagating out in the \ion{Fe}{12} 195.119  \AA\
monochromatic imaging, which is qualitatively consistent with the blue-shifts,
but no outward disturbances are detected in the \ion{Si}{7} 275.368 \AA\ line.
Time sequences in this line show downward propagating trends that are
consistent with the red-shifted emission of the line.

\begin{figure*}[htbp!]
\centering
\includegraphics[angle=90,width=17cm]{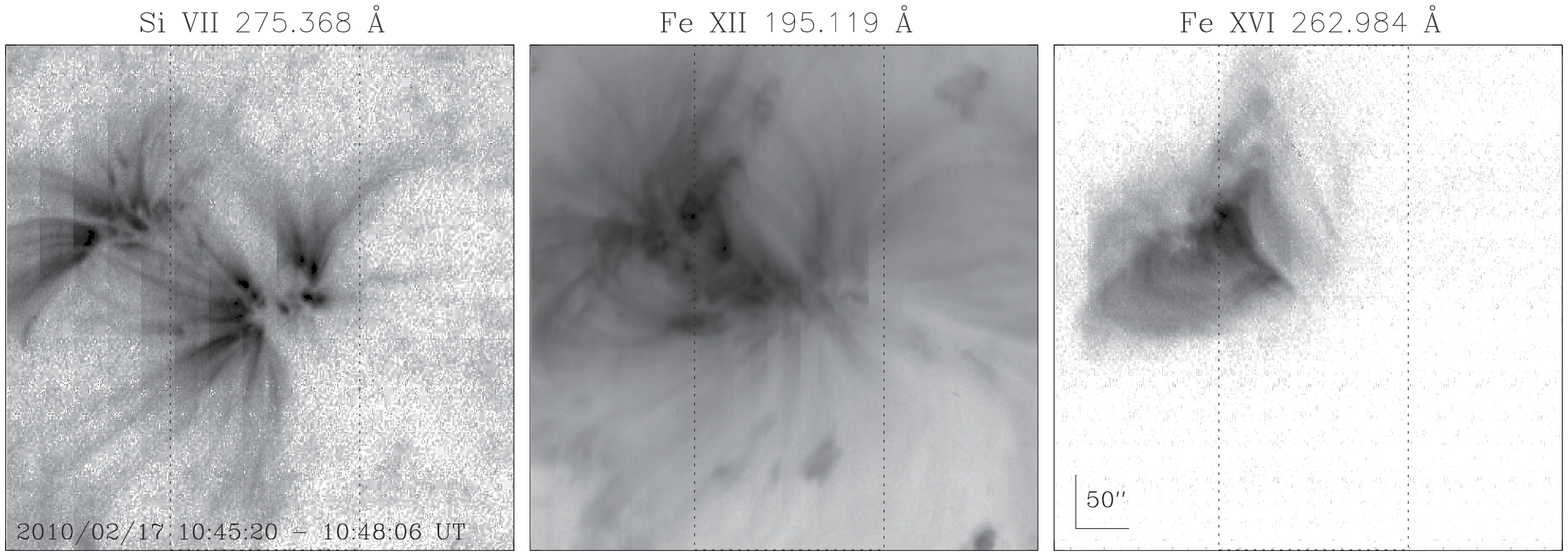}
\includegraphics[angle=90,width=17cm]{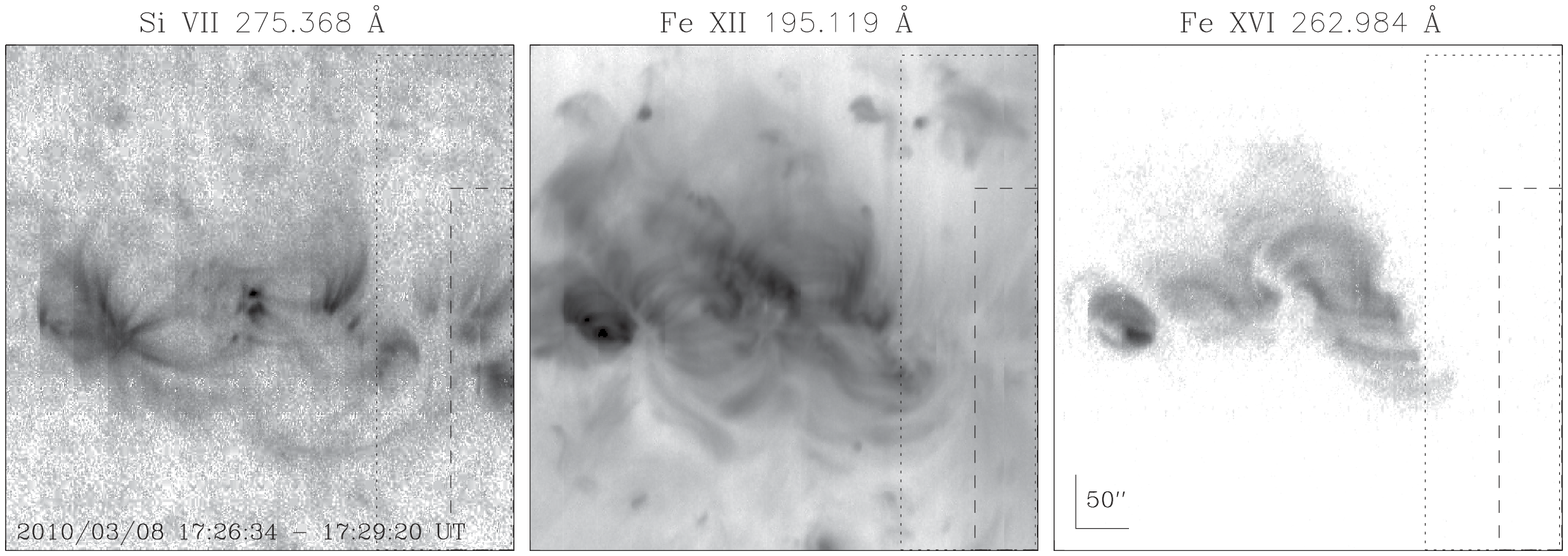}
\caption{EIS wide slit composite images in three spectral lines. Top: Active Region 11048 observed on 2010 February 2007. 
Bottom: Decaying Active Region 11045 observed on 2010 March 8. The dotted and dashed lines outline the field-of-view of 
the 1\arcsec\ slit rasters and the 2\arcsec\ fast scans respectively. The color scale is reversed: bright is black and dim
is white. The electronic version of the manuscript contains movies of both datasets.}
\label{fig:context_images}
\end{figure*}
\section{Observations}
We present results from two active region data sets obtained with
the Extreme-ultraviolet Imaging Spectrometer \citep{culhane2007}
on board {\it Hinode} \citep{kosugi2007}. The EIS instrument is a
high spatial (1\arcsec\ or 2\arcsec\ per pixel) and spectral (22 m\AA) 
resolution imaging spectrograph. It observes coronal and transition 
region spectral lines in two wavelength ranges: 170 -- 210 \AA\ and 
245 -- 290 \AA. Users can opt between narrow slit spectroscopy
(1\arcsec\ and 2\arcsec\ slits) or wide slit imaging (40\arcsec\ and
266\arcsec\ slits).

EIS observed Active Region (AR) 11048 on 2010 February 17. The observing 
sequence (10:45 -- 15:27 UT) consisted of  $480\arcsec\times488\arcsec$ 
images made out of fifteen consecutive 10\,s exposures at adjacent solar
positions, resulting in a 3 minute cadence. Despite the fact that the full 
spectral range is exposed on the detector, only a limited number of spectral 
windows is retrieved. We will discuss here in more detail images from 
two spectral lines: \ion{Si}{7} 275.368 \AA\ and \ion{Fe}{12} \AA\ 195.119 
\AA. The spectral purity of these images is 0.9 \AA. The imaging sequence
was preceded and followed by two narrow slit (1\arcsec) $178\arcsec\times512\arcsec$
rasters. The  rasters are sparsely sampled: the slit takes 3\arcsec\ steps 
between every exposure (50 s), which allows a faster scanning of the target. 
The rasters serve as a spectral diagnostics reference, in particular for the 
line-of-sight velocity. Top panels in Fig.~\ref{fig:context_images} show 
slot images of the active region for the two lines of interest, plus 
\ion{Fe}{16} 262.98 \AA. The dotted line encloses the field-of-view of the 
rasters. They are located over the low density \ion{Fe}{12} regions at the 
periphery of the active region. Notice that in \ion{Si}{7} that region shows 
also bright high density loops. Some of the spectral properties of this AR 
were also discussed by \citet{warren2010b}.

On 2010 March 8 EIS observed the decaying AR 11045. The observing
sequence starts with a context $128\arcsec\times512\arcsec$ 
fully sampled raster (1\arcsec\ slit, 60 s exposures), followed by the 
3 minute cadence slot imaging (17:23 -- 20:55 UT). To investigate the
short term variability of the spectral signatures, the sequence is 
followed by a set of sparsely sampled rasters (2\arcsec\ slit, 
4\arcsec\ steps) that cover a $58\arcsec\times368\arcsec$ area in 6 
minutes from 21:41 UT until 00:03 UT. The bottom panels of Fig.~\ref{fig:context_images} 
show a view of the active region and the different fields-of-view.
The electronic version of the manuscript contains movies of both 
datasets.

Data were processed using standard EIS software. This involves subtraction
of the dark current and correction of artifacts like cosmic rays and 
warm and hot pixels. Images were co-aligned using standard cross-correlation 
techniques. This removes both the spacecraft jitter and the displacement of 
the slot images along the wavelength direction on the detector due to the 
orbital changes in temperature \citep{brown2007}.

\begin{figure*}[htbp!]
\centering
\includegraphics[angle=90,height=15cm]{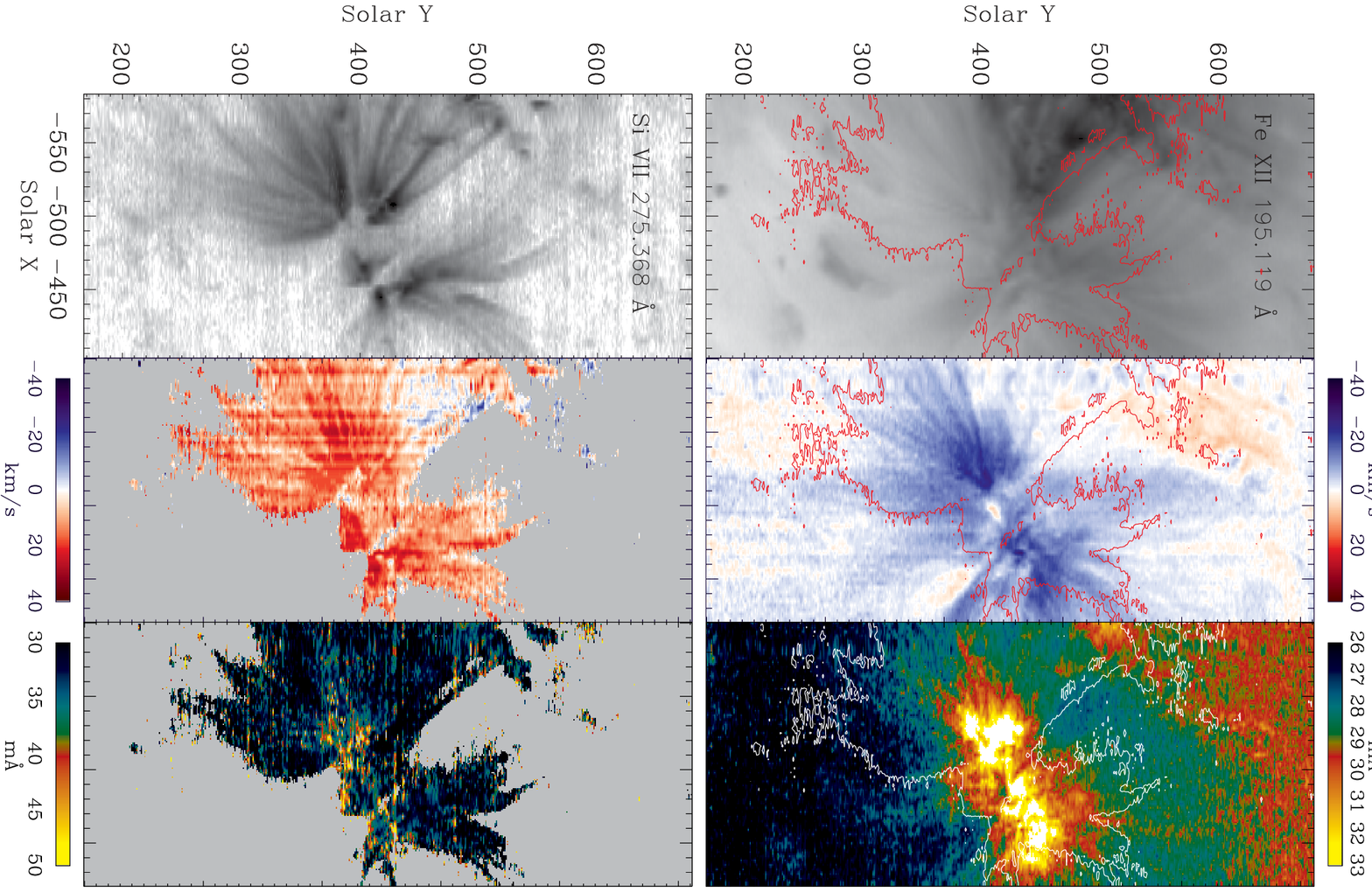}
\hspace{1cm}
\includegraphics[angle=90,height=15cm]{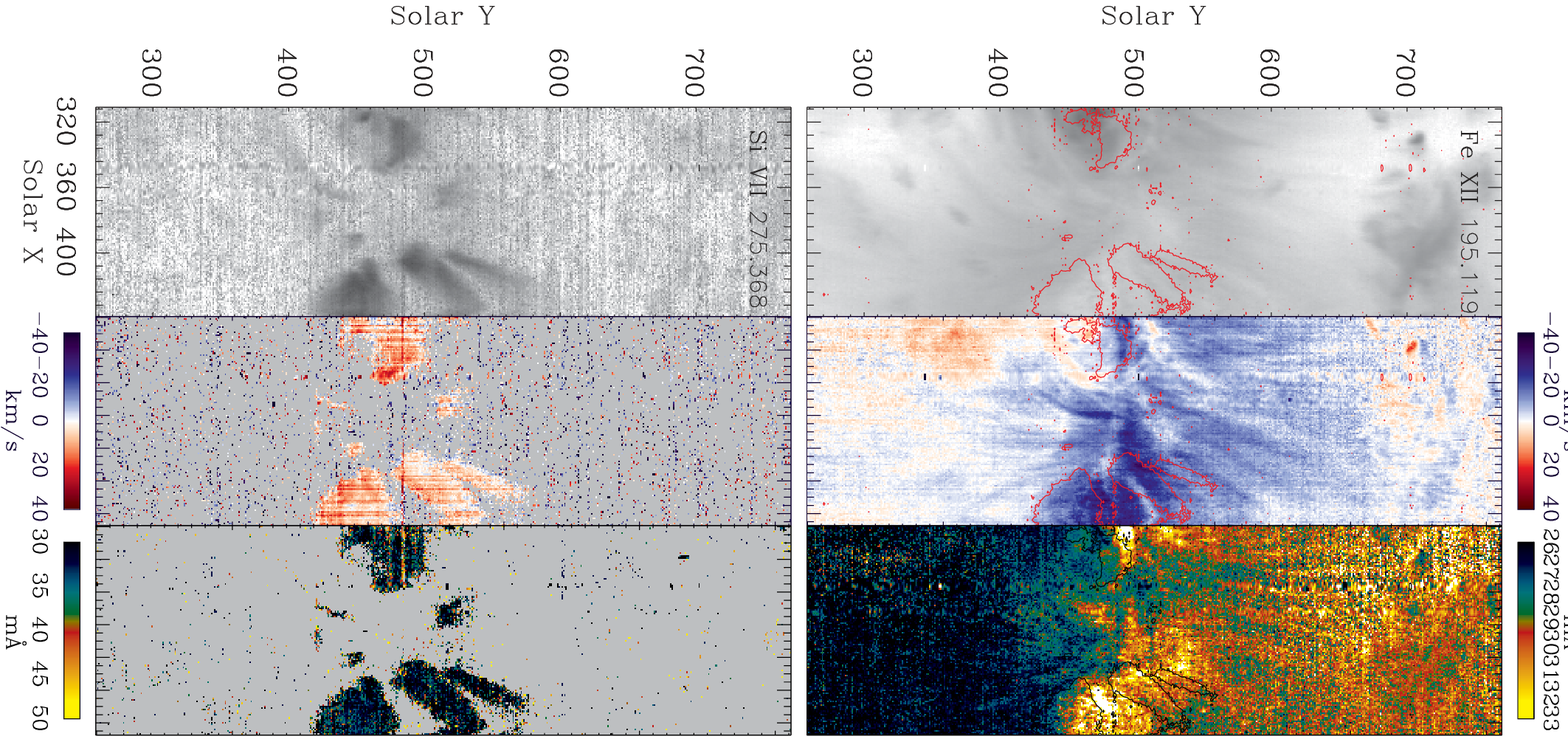}
\caption{EIS raster scans at the periphery of both active regions. Left: 2010 February 17 (04:25 -- 05:16 UT); actual raster is 
undersampled (1\arcsec\ slit 50 s exposures every 3\arcsec). Shown is the interpolated image along Solar X. Right:
2010 March 8 (15:02 -- 17:14 UT); fully sampled raster. Top row corresponds to \ion{Fe}{12} 195.119 \AA\ line and bottom to 
\ion{Si}{7} 275.368 \AA.  In each side, the left panels are radiances, middle panels are Doppler velocities and right panels are 
spectral widths. \ion{Si}{7} radiance contours over the \ion{Fe}{12} parameters for reference.}
\label{fig:rasters}
\end{figure*}

\section{Results}
Investigations of apparent motions in active region peripheral loops have 
been presented before. The novelty of this work is two-fold. Firstly, we
provide the first look at the motions with relatively high cadence (3 minute)
monochromatic (0.9 \AA) imaging, which should clarify any ambiguities from
line blending that is present  in EUV imagers. Secondly, we present the first
analysis of the short timescale spectral variability of the blue-shifted  
low density regions. Prior to these results we will discuss the morphological
relationship between the bright red-shifted \ion{Si}{7} loops and the dim 
blue-shifted \ion{Fe}{12} loops.

\subsection{Morphology}
Fig.~\ref{fig:rasters} shows the rastered areas of both active regions 
outlined with dotted lines in Fig.~\ref{fig:context_images}. The first column
is the radiance, followed by Doppler velocities and line widths. These 
quantities correspond to the Gaussian fit of the spectral profiles, a two 
Gaussian fit in the case of the \ion{Fe}{12} 195.119 \AA\ to account for 
the blend with the \ion{Fe}{12} 195.179 \AA\ line \citep{young2009}. Contours
of the \ion{Si}{7} emission have been put on top of the \ion{Fe}{12} maps
for reference. Note that the intensity scaling is the same as in Fig.~\ref{fig:context_images}.

There are several difficulties with measuring Doppler velocities with 
EIS and \citet{warren2010b} describe them in some detail. First, we need
to correct for the orbital drift of the spectrum on the detector and then
we have to assume a reference wavelength as our zero velocity. We use 195.119 
\AA\ and 275.368 \AA\ , as suggested by \citet{warren2010b}

The blue-shifts and width enhancements are located in regions that are dim 
compared to the core of the active region. This is now well known from 
EIS spectra \citep{doschek2008,harra2008,delzanna2008}. The \ion{Si}{7} 
line is weak outside the clearly defined loops, so the Doppler velocities 
and widths are only shown for the areas with high signal-to-noise ($\approx 7$). As 
discussed by \citet{warren2010b} these cool loops are consistently red-shifted. 
Width enhancements can also be detected in the March 8 dataset.  
They mostly correspond to the dim areas in between the bright loop footpoints.

\begin{figure*}[htbp!]
\centering
\includegraphics[angle=90,width=8.5cm]{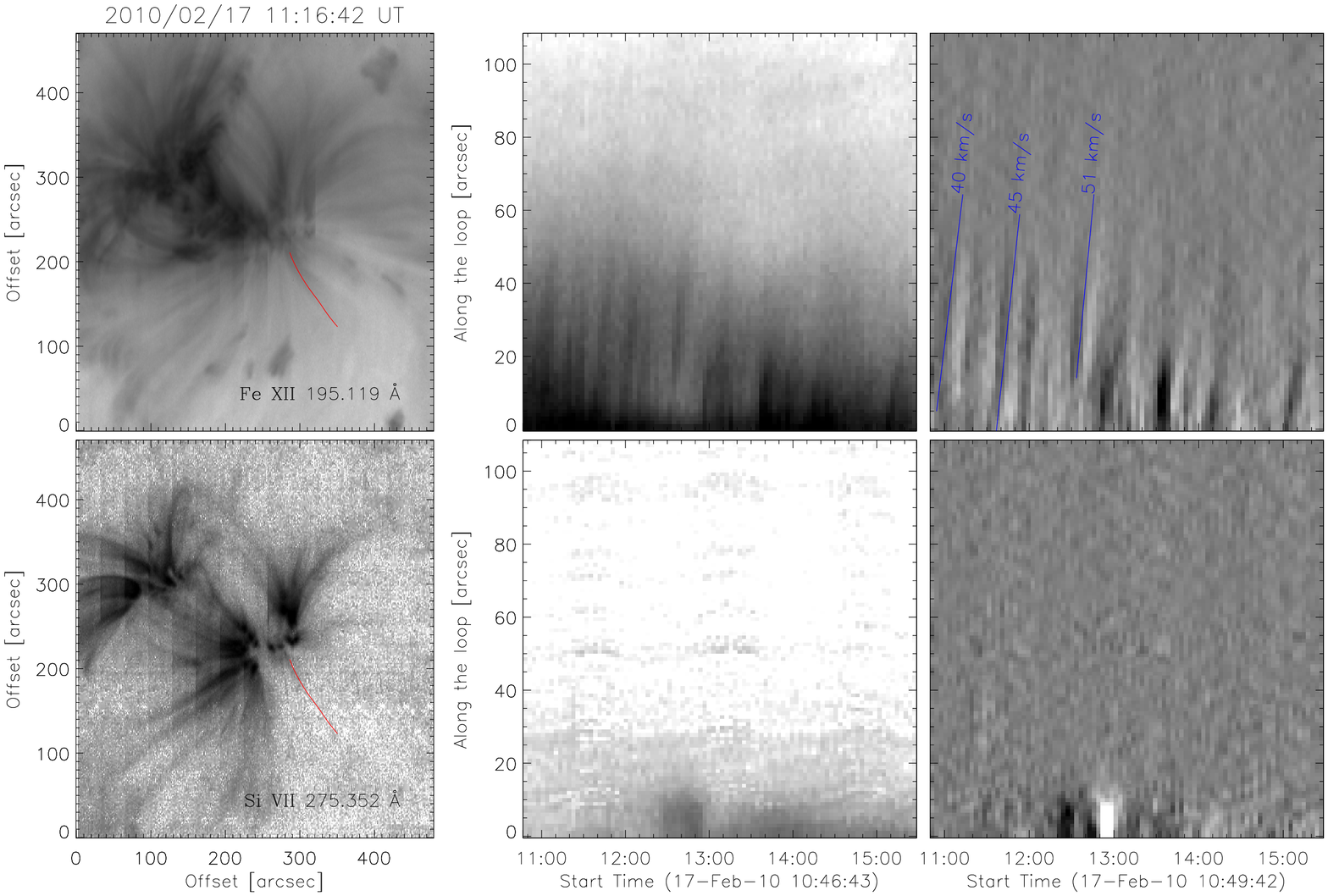}
\includegraphics[angle=90,width=8.5cm]{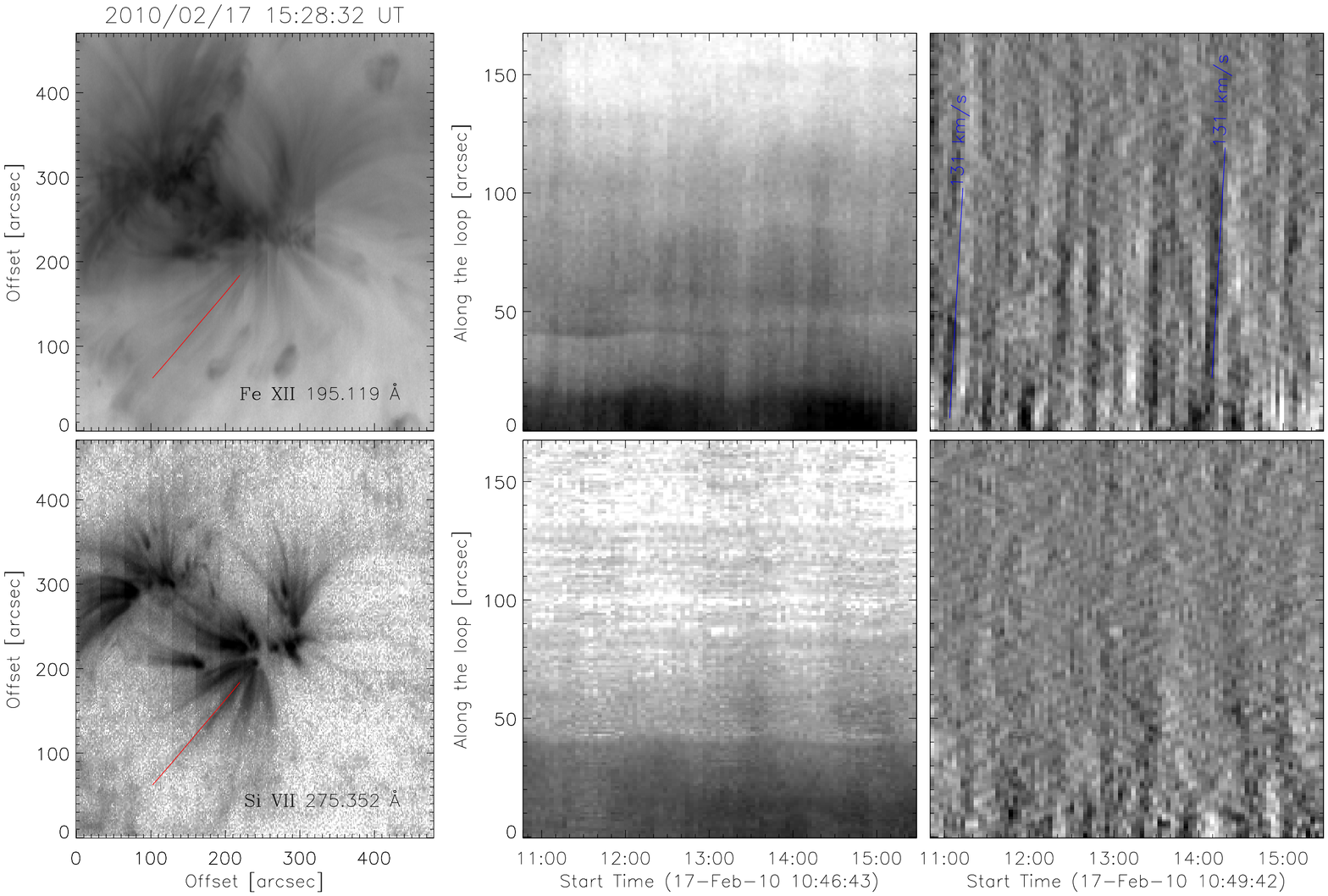}
\includegraphics[angle=90,width=8.5cm]{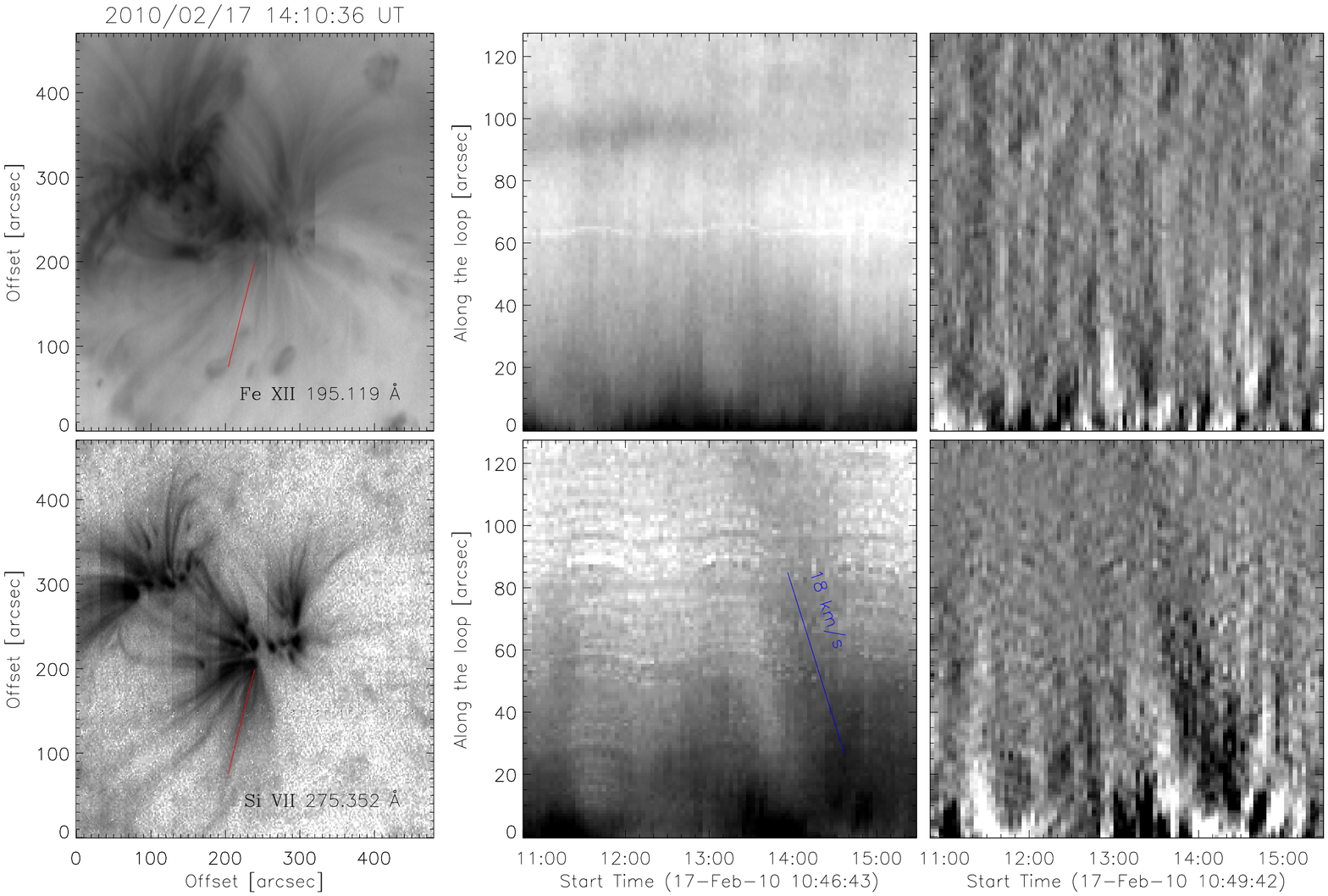}
\includegraphics[angle=90,width=8.5cm]{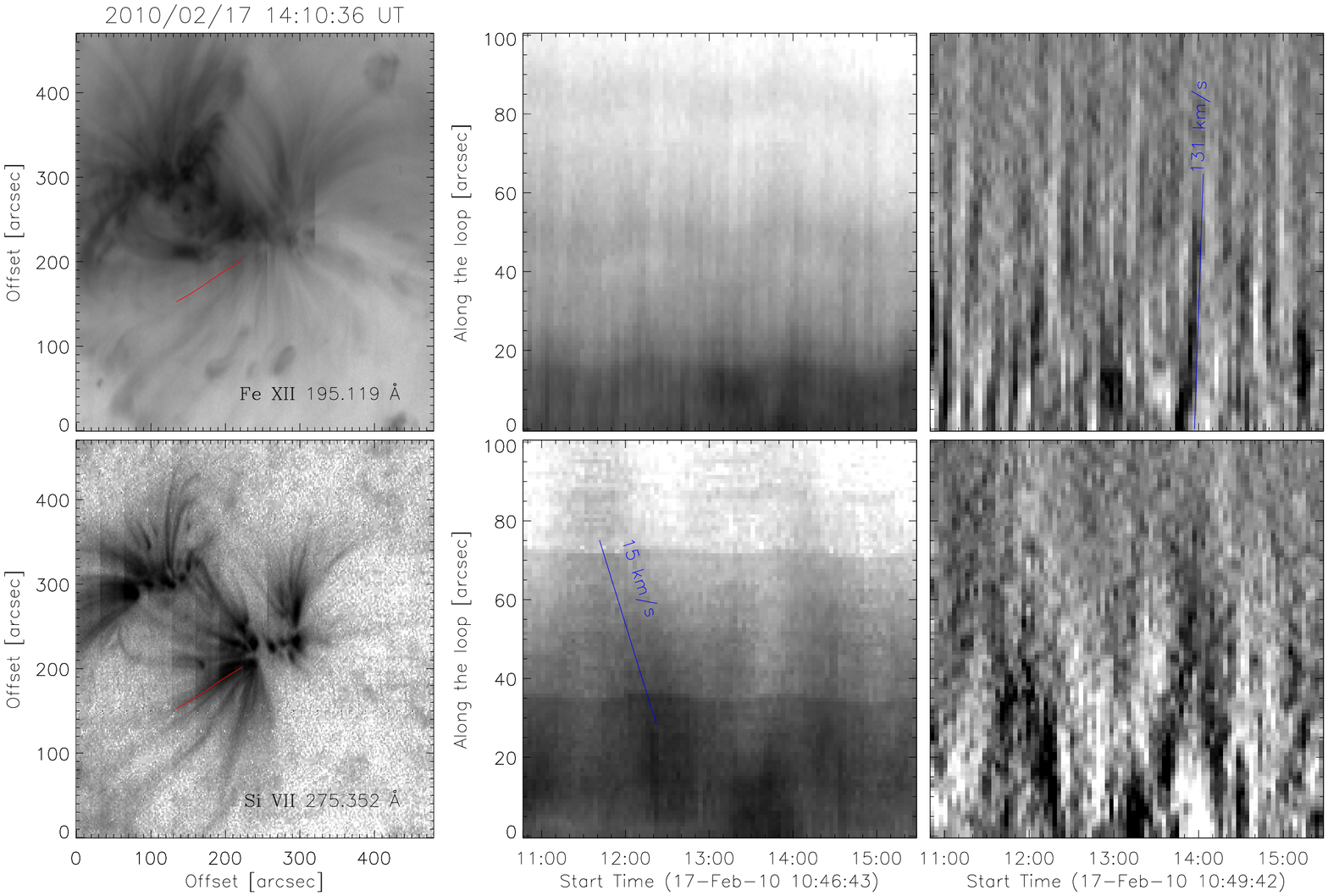}
\caption{Sample loops in the February 17 dataset. Each six panel display at each corner of the figure is a comparison of the 
\ion{Fe}{12} 195.119 \AA\ and \ion{Si}{7} 275.368 \AA\ intensity fluctuations along a loop structure outlined in red. 
The left-most panels shows the context slot image. The middle panel is the time-distance plot, namely the radiance changes 
as a function of time and position along the loop. The right-most panel is the running difference of the time-distance plot. }
\label{fig:rdiff1}
\end{figure*}

\begin{figure*}[htbp!]
\centering
\includegraphics[angle=90,width=8.5cm]{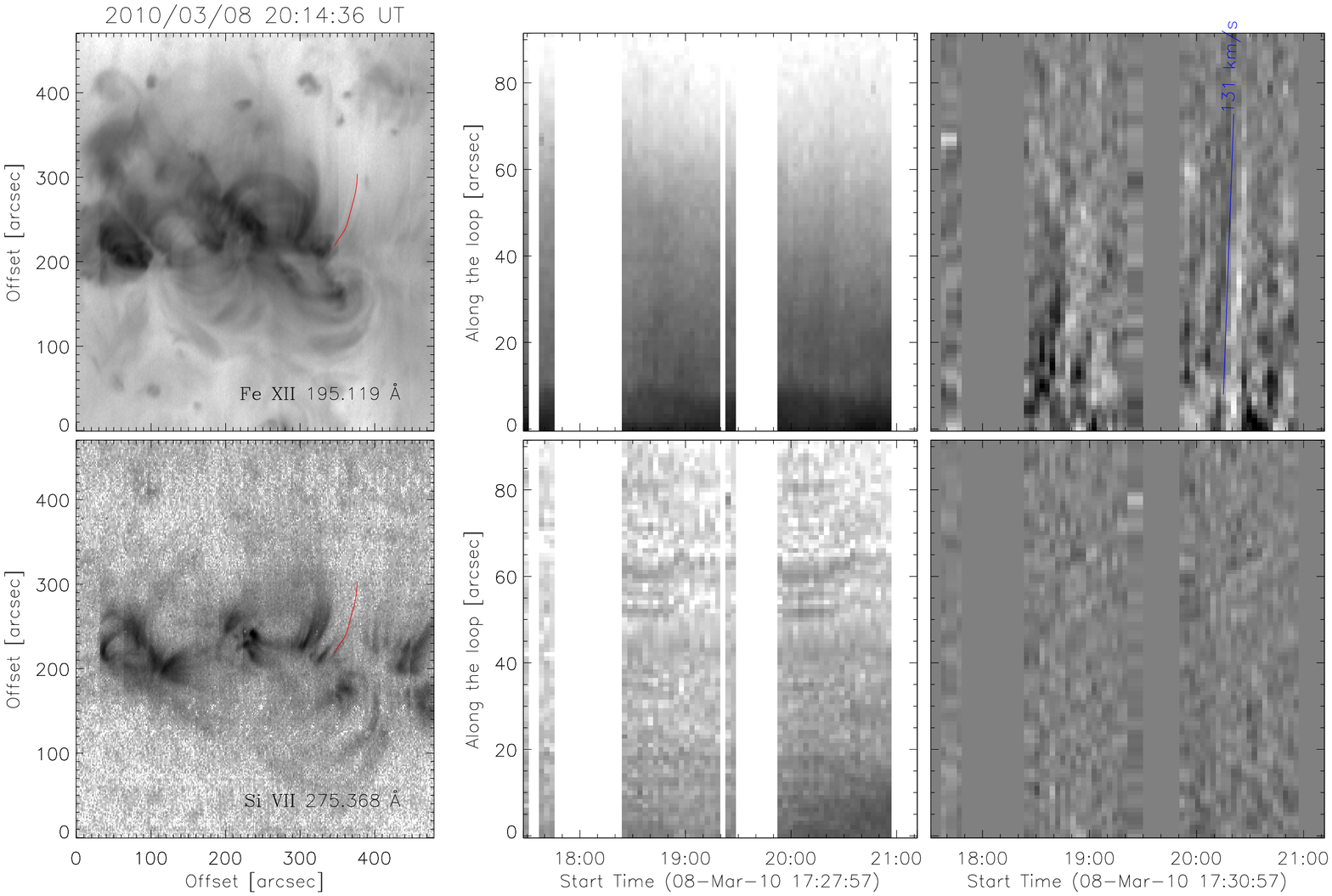}
\includegraphics[angle=90,width=8.5cm]{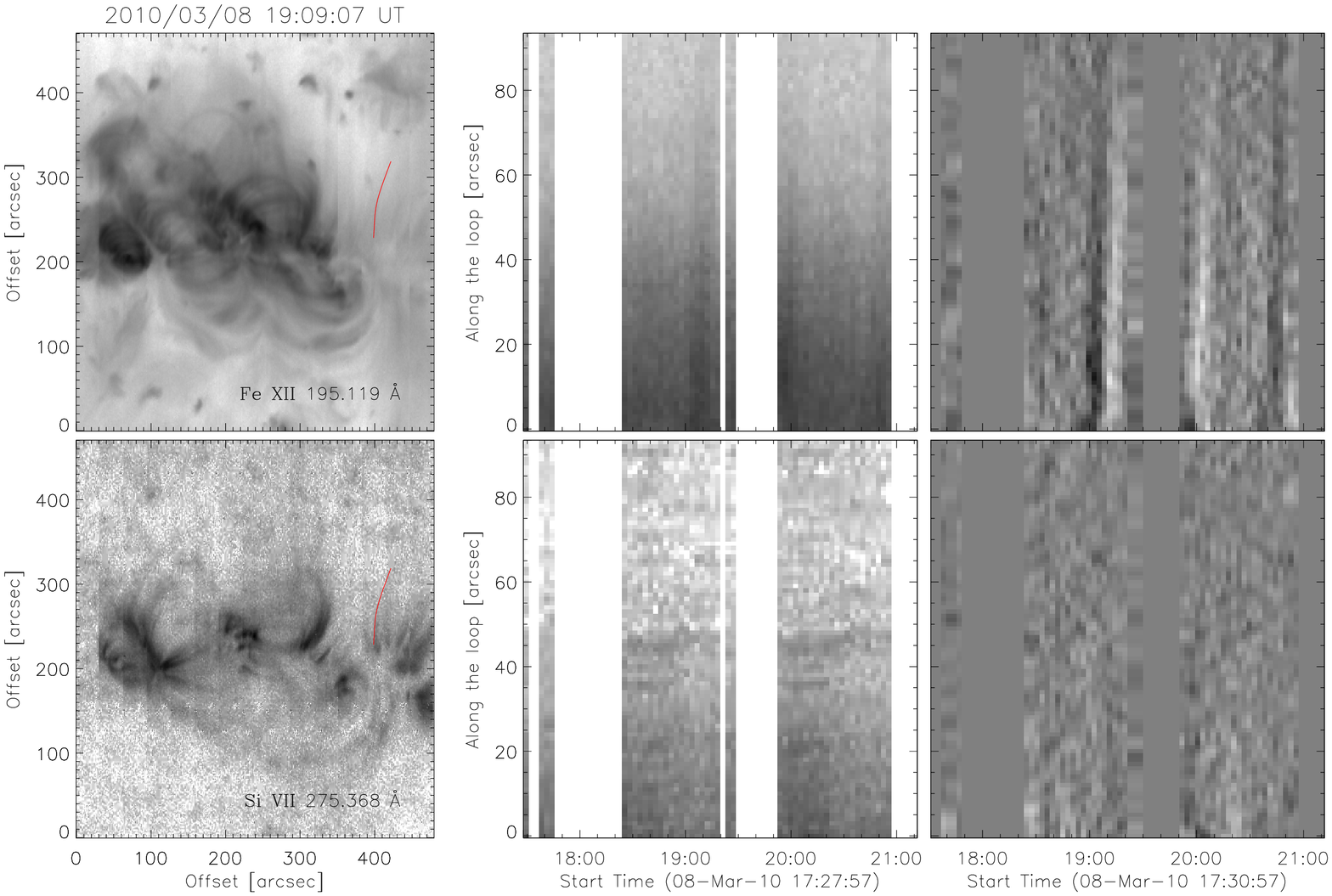}
\includegraphics[angle=90,width=8.5cm]{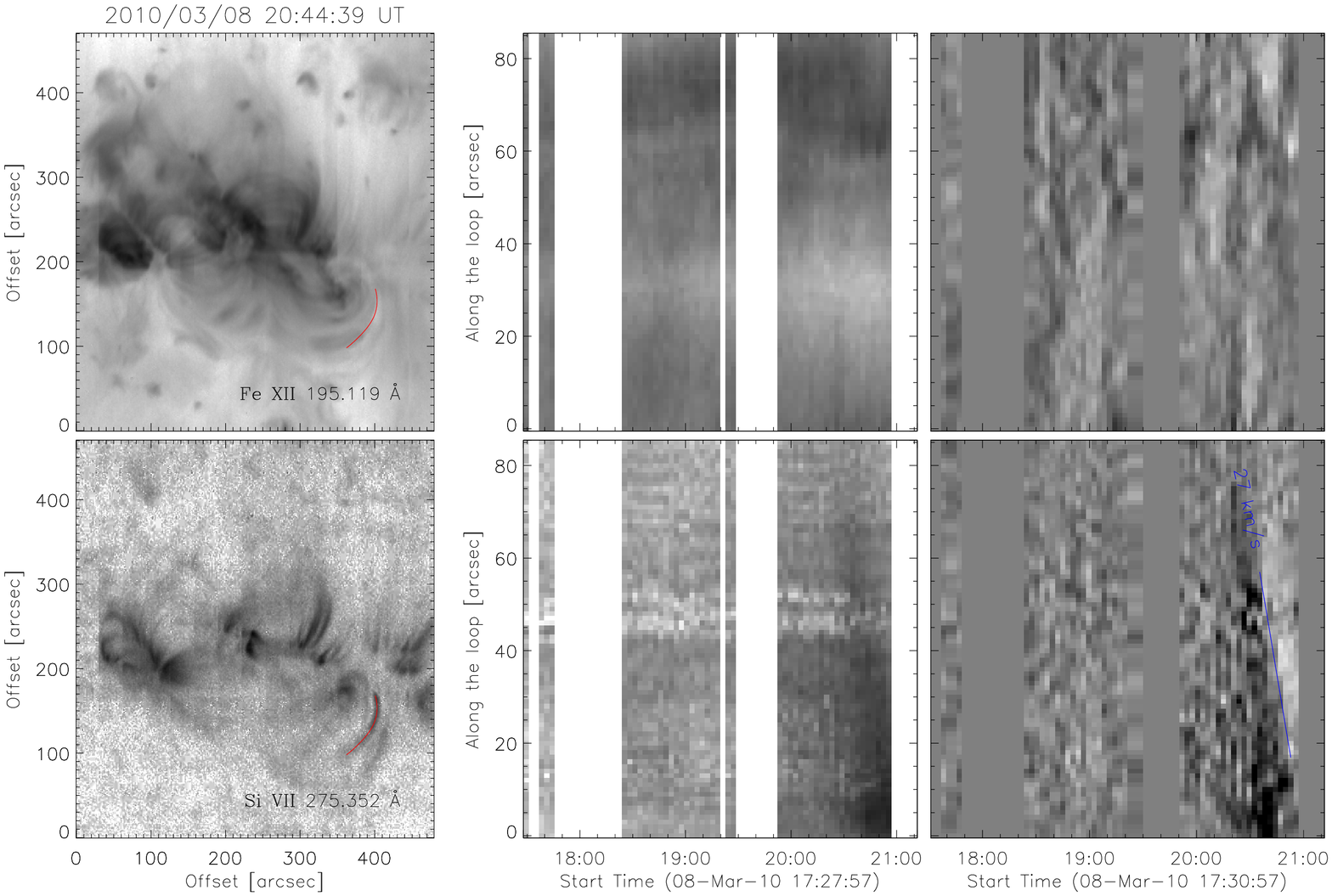}
\includegraphics[angle=90,width=8.5cm]{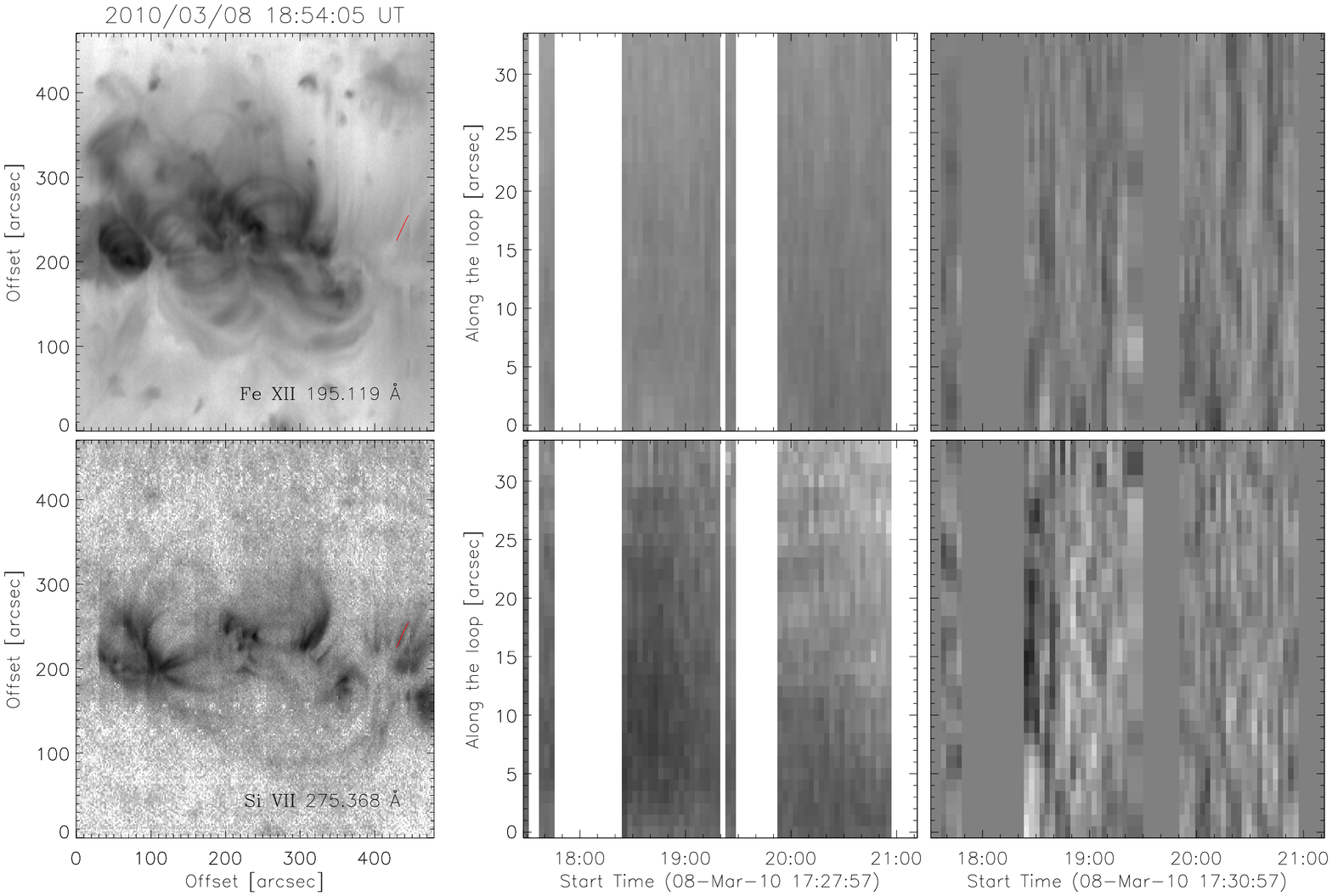}
\caption{Same as Fig.~\ref{fig:rdiff1} for the March 8 dataset. White bands represent data gaps.}
\label{fig:rdiff2}
\end{figure*}

In general, Fig.~\ref{fig:rasters} suggests that the bright \ion{Si}{7} 
loops are co-spatial with the \ion{Fe}{12} blue-shifts. There are differences,
however, when we look at the details. Extended loop structures in \ion{Fe}{12}
have a correspondence in \ion{Si}{7} in the February 17 dataset, but not as
much on March 8. Some of the \ion{Fe}{12} outward propagating motions, that 
will be discussed in the next section, occur in areas where no sizable \ion{Si}{7}
emission can be seen, for instance the South West corner of the February
17 dataset. An explanation could be that at this location only the footpoints 
of the loops hosting the disturbances have a cooler signature. That is what
Movie 1, available in the electronic version, shows. The March 8 dataset, however, 
shows significant areas of the field-of-view that have dim \ion{Fe}{12} emission 
and strong blue-shifts with no noticeable signatures in the cooler line, e.g., around 
[385\arcsec,490\arcsec]. Movie 2
confirms this and rules out the possibility of a time dependent origin for the 
discrepancy. What we can not completely rule out is that the absence of a
cooler plasma counterpart in these areas is just due to the instrument's 
sensitivity. 

\subsection{Time-dependent imaging}
Movie 1 and Movie 2, available in the electronic version of the manuscript, 
show the variability of the two active regions' peripheral loops
for the two spectral lines, \ion{Si}{7} 275.368 \AA\ and \ion{Fe}{12} 195.119 \AA, side 
by side. The February 17 dataset shows \ion{Fe}{12} disturbances propagating radially 
out from the blue-shifted area presented in Fig.~\ref{fig:rasters}. This also occurs in
the March 8 observations. These apparent motions are comparable to the ones observed with
{\it TRACE} and XRT/{\it Hinode} \citep{mcintosh2009,sakao2007}. Our observations therefore
confirm that the phenomenon is observable at 1\,MK: the EIS wide slit images isolate a very
narrow (0.9 \AA) and blend free spectral region around the line \citep[e.g.,][]{ugarte-urra2009}. 
{\it TRACE} 171 \AA\ and 
195 \AA\ passbands can be ambiguous in this respect because both bands  have significant 
contributions from plasma at 0.6 MK, namely \ion{Fe}{8} and \ion{Fe}{9} lines 
\citep{delzanna2003b}, precisely the lines in which the cool peripheral loops manifest 
themselves. An analogous argument can be made about the broad temperature response of XRT 
filters. 

These outward propagating disturbances are not observed in the \ion{Si}{7} images.
Fig.~\ref{fig:rdiff1} and Fig.~\ref{fig:rdiff2} show examples of time distance plots
for various representative locations in both active regions. Each figure compares the 
time dependent intensity fluctuations of \ion{Fe}{12} and \ion{Si}{7} along four loop 
segments. The time-distance plots were constructed in the 
following manner. Following \citet{warren2010} we first manually selected the points 
along the segment. The points were used as spline knots to define a loop coordinate
system ($s$,$t$) where $s$ goes along the loop's axis and $t$ perpendicular to it. 
From this interpolated straightened loop segment we extracted
the intensity along the axis (1\arcsec\ across) and plotted it as function of time. This
is shown in logarithmic scale in the middle panel. To its right we also show the 
running difference, i.e. the difference between two consecutive intensity values at
a particular location. The running difference time-distance plots have been smoothed 
(boxcar of three pixels) to increase the signal-to-noise. 

The loops are shown as a solid red line. The top two loops were selected based on the
propagating lanes followed by the disturbances. The bottom two outline clearly defined
loops in the \ion{Si}{7} images. The time-distance plots in \ion{Fe}{12} show features
very similar to ones seen by {\it TRACE} and XRT: recurrent inclined ridges that
represent intensity fluctuations propagating along the loops as function of time. The
velocity of propagation ranges between 40 --  130$\rm \,km\,s^{-1}$, as the reference
solid blue lines show. These are propagation velocities on the plane-of-the-sky and 
therefore only lower limits. Therefore, for \ion{Fe}{12}, we find qualitative consistency 
between the spectral blue-shifts and the upward apparent motions in the imaging.

The \ion{Si}{7} time-distance plots of the top two loops do not exhibit any distinct
variability. In particular, most of the emission of this line at the South West corner of 
the image's field-of-view is noise. The bottom loops do experience some variability. We 
observe downward propagating intensity changes at lower speeds 15 -- 20$\rm \,km\,s^{-1}$, 
qualitatively consistent with the spectral redshifts measured in the rasters. Similar 
downflows in \ion{Si}{7} loops have also been observed off-limb \citep{ugarte-urra2009}.
If this is a manifestation of cooling loops, we do not see any evidence of them in 
the hotter line.
The picture, however, can be more complicated than just downflows, as the last panel in 
Fig.~\ref{fig:rdiff1} shows, where the ridges show oppositely directed trends. 
\citet{schrijver1999} already pointed out the rather complex evolution of these fan
loops comparing them to ``rippling curtains''. Ultimately, what seems clear from the
comparison of the time-distance plots is that there is no evidence that the structures and
the intensity fluctuations they exhibit in \ion{Fe}{12} (1.3 MK) and \ion{Si}{7} (0.6 MK) are 
related to one another. As stated before, sensitivity can be an issue in the case of the
cooler and weaker line, however, it does not explain why we do not see the \ion{Si}{7} 
downward trends in \ion{Fe}{12}.

The time-distance plots of the March 8 dataset may appear less compelling due to the 
data gaps. Movies 1 and Movie 2 show, however, that both examples portray the same 
phenomenom. This is important, because the March 8 dataset allows us to look into the
time-dependent spectroscopic properties.

\begin{figure*}[htbp!]
\centering
\includegraphics[]{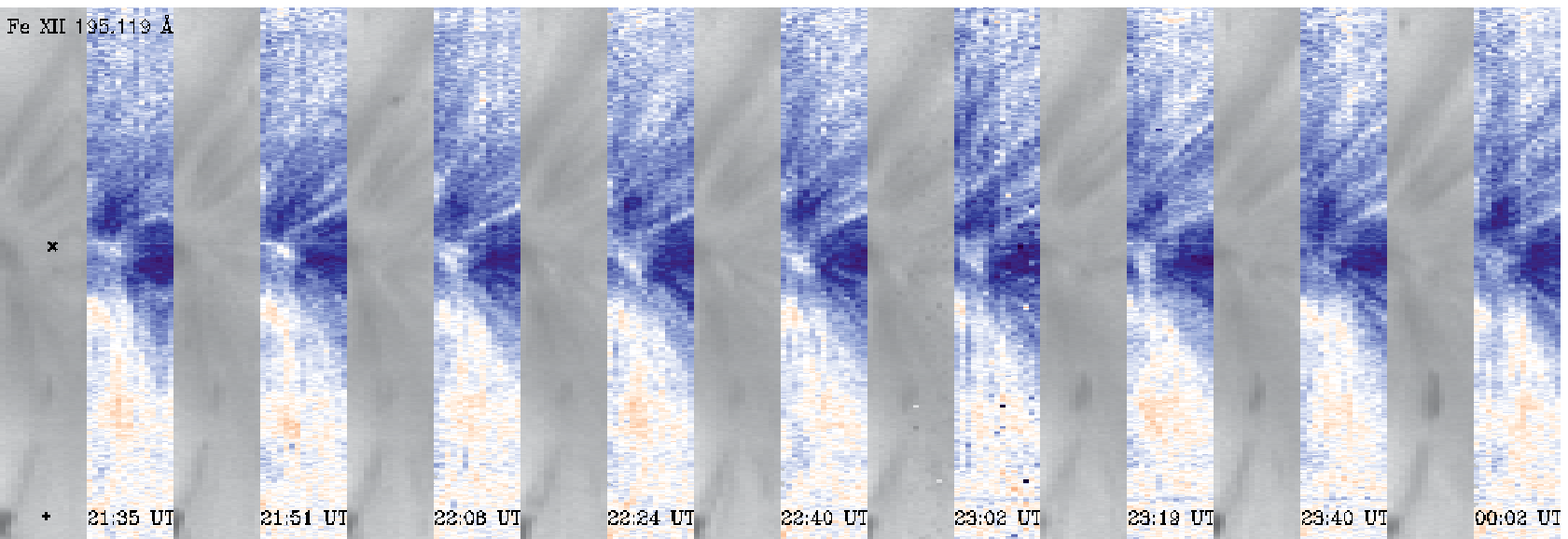}
\includegraphics[]{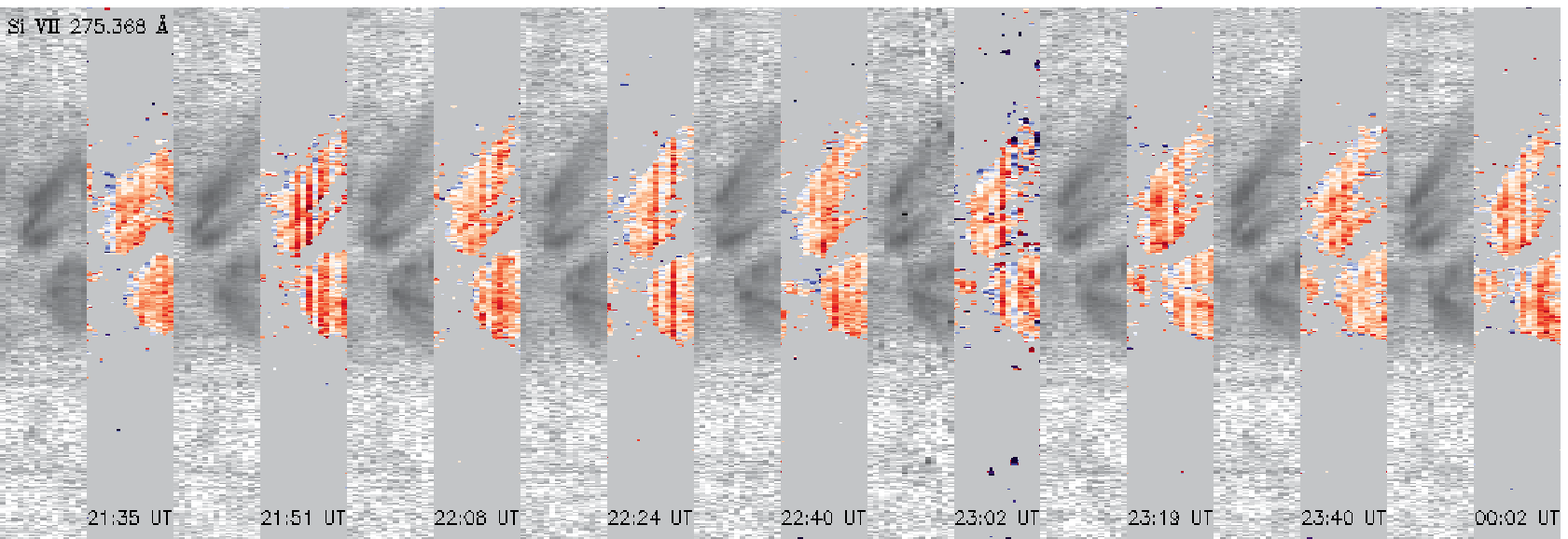}
\caption{Fast scans (2\arcsec\ slit) showing the radiance and Doppler shift changes as a function of time in the March 8 
dataset. Field of view is $60\arcsec\times368\arcsec$. Observing sequence only samples 2\arcsec\ in every 6\arcsec\  
(4\arcsec\ steps between exposures). For presentation purposes, we interpolate in between. Only shown one third of the frames.
The cross and plus symbols are references for the spectral profiles in Fig.~\ref{fig:spec_profiles}.}
\label{fig:scan_steps}
\end{figure*}
\begin{figure*}[htbp!]
\centering
\includegraphics[width=17cm]{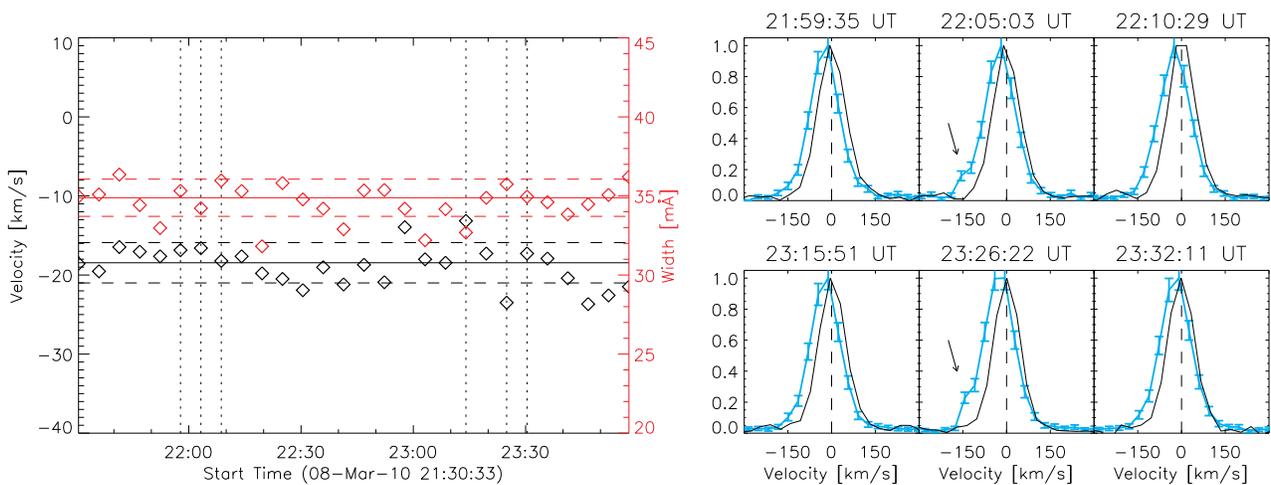}
\caption{Left: Doppler velocity and width of the \ion{Fe}{12} 195.119 \AA\ line
as a function of time for a representative location in the fast scans (cross symbol
in Fig.~\ref{fig:scan_steps}). The
average value and the standard deviation from it are represented by the horizontal
solid and dotted lines. Right: spectral line profiles for six different times 
(vertical dotted lines) in blue. In black is shown a reference profile for a location
outside of the blue-shifted region (plus symbol in Fig.~\ref{fig:scan_steps}).}
\label{fig:spec_profiles}
\end{figure*}

\subsection{Time-dependent spectroscopy}
We have confirmed that there are outward propagating disturbances at 1 MK
that originate in blue shifted areas at the periphery of active regions.
The time-distance plots clearly show that the perturbations are
discrete in nature, but spectroscopically we only know that they can 
persist for days \citep{bryans2010}.

We investigate therefore the variability of the Doppler shifts in 
short timescales to determine whether that transient nature is also
reflected in the line profiles. Fig.~\ref{fig:scan_steps} shows
a sequence of radiances and Doppler shifts for the two spectral 
lines in the March 8 dataset. The field-of-view corresponds to the
dashed line in Fig.~\ref{fig:context_images}. The rasters, consisting
of fifteen 20 s exposures with 4\arcsec\ steps in between, result in 
a 5 minute cadence. We only show a third of that cadence in the figure.
It confirms that \ion{Fe}{12} and \ion{Si}{7} are consistently 
blue and red shifted. 

Fig.~\ref{fig:spec_profiles} shows in its left panel the Doppler shift and 
width variations of the \ion{Fe}{12} line at  one particular location. 
While the changes observed in the main component of the line are small, 
within the standard deviation (dashed lines), we do detect 
asymmetric wing enhancements in the blue side of the line 
profiles, on timescales as short as the 5 minute cadence (see right panels 
in Fig.~\ref{fig:spec_profiles}). The enhancements occur at velocities 
of $\approx150\,\rm km\,s^{-1}$. This result confirms the suspicion that 
the reported asymmetric profiles discussed by \citet{bryans2010} and 
\citet{mcintosh2009} can have a short term nature, which favors an
association with the discrete outward propagating intensity disturbances.
The spectral line, however, is dominated by emission that exhibits a rather 
constant line-of-sight velocity of $\approx20\,\rm km\,s^{-1}$.

\section{Conclusions}
We have investigated the time dependent spectral properties of areas
at the periphery of active region cores. In \ion{Fe}{12} 195.119 \AA\ 
(1.3 MK) we find recurrent intensity disturbances originating in low density regions
that propagate outwards along loop structures at projected velocities of 40 -- 130 
$\rm \,km\,s^{-1}$. This is consistent with the characteristic blue-shifted
emission measured along the line-of-sight at their footpoints. The spectral 
profiles are dominated by small and constant shifts ($\approx20\rm \,km\,s^{-1}$) 
in the main component of the spectral line, but reveal transient intensity enhancements in 
the blue wing at velocities of $\approx150\,\rm km^{-1}$, on timescales as short 
as the available cadence: 5 minutes.
This is supportive of the interpretations of the fluctuations as a result
of transient events \citep{mcintosh2009}, although we have not been 
able to establish a connection with the cooler temperatures.

In \ion{Si}{7} 275.368 \AA\ (0.6 MK) the same areas alternate apparent voids
with high density bright loops characterized by emission which is consistently
red-shifted. No sign of the recurrent outward propagating disturbances is
detected, but when trends are present they are generally downward at velocities of
$ \approx15 - 20\rm \,km\,s^{-1}$, qualitatively consistent with the spectral
measurements at the footpoints. We find therefore no evidence that the structures 
and the intensity fluctuations visible at 1 MK and 0.6 MK are inter-related.

Given that {\it TRACE} 195 \AA\ images can be contaminated by the emission of 0.6 MK 
cool lines \citep{delzanna2003b}, this is the first confirmation that the outward 
propagating disturbances take place at 1 MK temperatures. Outward 
propagating perturbations have also been observed by {\it TRACE} in its 
171 \AA\ filter images, which are mostly dominated by emission from 
\ion{Fe}{9}-{\small\rmfamily X} lines. We do not observe those perturbations 
in the \ion{Si}{7} fan loops, which suggests either that the EIS instrument is 
not sensitive enough to detect them or that the perturbations observed on the 
171 \AA\ fan loops are coronal in nature and a result of the broader temperature 
response. New coordinated observations of EIS and AIA/{\it SDO}, with
similar filters, should shed some light on this issue.

\acknowledgments The authors acknowledge support from the NASA {\it Hinode} program.
{\it Hinode} is a Japanese mission developed and launched by ISAS/JAXA, with
NAOJ as domestic partner and NASA and STFC (UK) as international partners. It is operated
by these agencies in co-operation with ESA and NSC (Norway).


\end{document}